\documentclass[%
 aip,
 amsmath,amssymb,
reprint, onecolumn 
]{revtex4-1}

\usepackage{graphicx}
\usepackage{dcolumn}
\usepackage{bm}

\usepackage[utf8]{inputenc}
\usepackage[T1]{fontenc}
\usepackage{mathptmx}
\usepackage{etoolbox}
\usepackage{color}

\newcommand\mnras{{MNRAS}}     
\newcommand\aap{{A\&A}}        
\newcommand\pasj{{PASJ}}       
\newcommand\apjl{{ApJ}}        
\newcommand\apjs{{ApJS}}       

\newcommand\araa{{Ann Rev A Ap}}
\newcommand\solphys{{Sol Phys}}

\newcommand*\jgr{J. Geophys. Res}
\newcommand*\aaps{A\&AS}
\newcommand*\rmxaa{Revista Mexicana de Astronomía y Astrofísica}

\makeatletter
\def\@email#1#2{%
 \endgroup
 \patchcmd{\titleblock@produce}
  {\frontmatter@RRAPformat}
  {\frontmatter@RRAPformat{\produce@RRAP{*#1\href{mailto:#2}{#2}}}\frontmatter@RRAPformat}
  {}{}
}%
\makeatother
\begin{document}

\preprint{AIP/123-QED}

\title[Temperature-reducing Shocks in radiative MHD]{Temperature-reducing shocks in optically-thin radiative MHD - analytical and numerical results}
\author{B. Snow}
 \email{b.snow@exeter.ac.uk}
\affiliation{ 
University of Exeter
}%

\date{\today}

\begin{abstract}
Shocks are often invoked as heating mechanisms in astrophysical systems, with both adiabatic compression and dissipative heating that leading to temperature increases.   
Whilst shocks are reasonably well understood for ideal magnetohydrodynamic (MHD) systems, in many astrophysical plasmas, radiation is an important phenomena, which can allow energy to leave the system. As such, energy becomes non-conservative which can fundamentally change the behaviour of shocks. The energy emitted through optically-thin radiation post-shock can exceed the thermal energy increase, resulting in shocks that reduce the temperature of the medium, i.e., cooling shocks that have a net decrease in temperature across the interface. 
In this paper, semi-analytical solutions for radiative shocks are derived to demonstrate that both cooling (temperature decreasing) and heating (temperature increasing) shock solutions are possible in radiative MHD. Numerical simulations of magnetic reconnection with optically-thin radiative losses also yield both heating and cooling shocks in roughly equal abundances. The detected cooling shocks feature a significantly lower pressure jump across the shock than their heating counterparts. The compression at the shock front leads to locally-enhanced radiative losses, resulting in significant cooling within a few grid cells in the upstream and downstream directions. 
The presence of temperature-reducing (cooling) shocks is critical in determining the thermal evolution, and heating or cooling,   across a wealth of radiative astrophysical plasmas.
\end{abstract}

\maketitle

\section{Introduction}
\label{sec:intro}

Shocks are a universal feature of astrophysical systems, for example supernovae \cite{Weaver1976,Chen2022}, bow shocks in stellar wind \cite{Wilkin1996}, accretion disk around black holes \cite{Das2002}, magnetic-reconnection driven flares \citep{Petschek1964}, umbral flashes above solar sunspots \cite{Beckers1969}, to name a few. 
Typically, shocks are considered to be near-discontinuous jumps in properties of the medium, and are intrinsically linked to heating though adiabatic compression of the medium and non-adiabatic dissipation (which is efficient due to the sharp interface between the upstream and downstream media). As such, shocks are often studied for their role in energy redistribution and heating of astrophysical systems.

In ideal (non-radiative) magnetohydrodynamic (MHD) systems, the role of shocks is reasonably well understood with analytical solution that demonstrate the necessity for temperature increases across the shock \cite{Goedbloed2010}. 
However, often astrophysical systems require extensions to the ideal MHD model to accurately capture the salient physics. A common extension of the MHD equations is radiative MHD, which uses a temperature-dependent loss function to approximate the energy radiated from an optically-thin plasma by photons. The optically-thin radiative MHD model has been widely applied to study the solar atmosphere and astrophysical systems  \citep{Gudiksen2011,Forbes1991,Avara2016,Bingert2011,Ni2015,Rozyczka1985}. The exact nature of the loss function depends on the abundances and relevant processes, and even the numerical techniques used to calculate the loss function \cite{Hitomi2018}.

The simplest inclusion of optically-thin radiative losses involves adding a temperature and density dependent source term to the energy equation. However, the inclusion of source terms in the underlying equations allows for significant departures from the ideal MHD for stable solutions to shocks. For shocks in the low-temperature ($T\lessapprox 10^4$K) interstellar medium, molecular disassociation is known to be an import process that can lead to strong cooling and effectively isothermal shocks \citep{Draine1983,Draine1993}. Radiative cooling has been shown to important in determining the temperature profile of stellar shocks \citep{Fokin2004}, producing less heating than would be predicted by MHD alone. Many analytical studies have been performed to determine the structure and temperature profile of radiative shocks finding weakly heating and iosthermal solutions \citep{Weaver1976,Das2002,Lesaffre2004}. However, an additional, unstudied solution class exists whereby the radiative losses within the shock exceed the adiabatic heating leading to shocks that have net temperature reductions in the postshock region, i.e., shocks that cool the plasma. In this paper, such shocks that have a net decrease in temperature across the interface are referred to as 'cooling shocks'.

This paper demonstrates the existence of cooling solutions for shocks in radiative MHD using realistic radiative loss curves for optically-thin losses in solar and astrophysical systems. Firstly, the theory is presented, along with semi-analytic solutions to the radiative MHD equations. Shocks are then analysed in a numerical simulation of magnetic reconnection that is tearing unstable. Shocks are detected using the ShockID methodology \citep{Snow2021ShockID} and the local properties perpendicular to the shock front analysed to identify both heating and cooling shock solutions. The presence of cooling shocks means that the thermal contribution of shocks in radiative plasmas requires careful consideration.

\section{Methods}

The underlying equations for a compressible, radiative magnetohydrodynamic (MHD) plasma are given by: 
\begin{gather}
\frac{\partial \rho}{\partial t} + \nabla \cdot (\rho \textbf{v}) = 0, \\
\frac{\partial}{\partial t} (\rho \textbf{v})+ \nabla \cdot \left( \rho \textbf{v} \textbf{v} + P \textbf{I} - \textbf{B B} + \frac{\textbf{B}^2}{2} \textbf{I} \right) = 0,\\
\frac{\partial}{\partial t} \left( e + \frac{\textbf{B}^2}{2} \right) + \nabla \cdot \left[ \textbf{v} ( e + P) -  (\textbf{v} \times \textbf{B}) \times \textbf{B} \right]  =  - \rho ^2 \Lambda(T) + \Phi_{H}, \\
\frac{\partial \textbf{B}}{\partial t} - \nabla \times (\textbf{v} \times \textbf{B}) = 0, \\
\nabla \cdot \textbf{B} =0, \\
e = \frac{P}{\gamma -1} + \frac{1}{2} \rho v ^2, 
\end{gather}
for density $\rho$, velocity $\textbf{v}=\left[ v_x,v_y,v_z \right]$,  magnetic field $\textbf{B}=\left[ B_x,B_y,B_z \right]$ and pressure $P$. The temperature is assumed to obey the ideal gas law $T=\frac{\gamma P}{\rho}$. Optically thin radiative losses are assumed to be of the form $\rho^2 \Lambda (T)$, where $\Lambda(T)$ is a temperature dependent function. The exact form of $\Lambda(T)$ often depends on the abundances of atomic/molecular species, and the relevant processes that result in photon emission.
Included in the equations is a heating term $\Phi_H$ which is included as an ad-hoc balance to the radiative losses such that equilibrium can exist initially. The heating term $\Phi_H$ is usually assumed to be constant in time since the heating processes typically occur on far longer timescales than the radiative cooling. In this paper, $\Phi_H$ will be assumed to be constant and uniform.

\subsection{Rankine-Hugoniot jump conditions (no losses)}

In this section, the analytic shock jump conditions are presented for an ideal plasma (with no radiatve losses). For the MHD equations, the shock jump conditions can be derived by considering the equations upstream ($^u$) and downstream ($^d$) of the shock in the deHoffmann-Teller shock frame, where the electric field is identically zero. The conservative nature of the MHD equations means that for a steady-state shock in the deHoffmann-Teller frame, mass, momentum and energy can be equated sufficiently upstream and downstream. The MHD equations have no source terms and hence can be integrated across the shock to become:  
\begin{gather}
    \left[\rho v_x  \right]^u _d = 0,  \\
    \left[\rho v_x^2 +P +\frac{B_y^2}{2} \right]^u _d = 0, \\
    \left[\rho v_x v_y -B_x B_y \right]^u _d = 0, \\
    \left[ v_{x} \left( \frac{\gamma}{\gamma -1} P + \frac{1}{2} \rho v^2 \right) \right]^u _d =0, \label{Eqn:EnergyJump} \\ 
    \left[B_x \right]^u _d = 0, \\
    \left[v_x B_y -v_y B_x   \right]^u _d = 0, 
\end{gather}
where this notation means that
\begin{gather}
    \left[ Q \right]^u _d \equiv Q^u - Q^d,
\end{gather}
for any quantity $Q$ evaluated $^u$ upstream (pre-shock) or $^d$ downstream (post-shock). Shocks are essentially 1D structures since the gradients are greatest in one direction and hence the jump conditions can be given in terms of along the shock direction (here this is the $x$-direction) or perpendicular to it ($y-$direction). 

From these equations, the jump conditions for primitive plasma properties can be derived\cite{Goedbloed2010,Snow2021} as:
\begin{gather}
    \frac{\rho^d}{\rho^u}=r \\
    \frac{v_x^d}{v_x^u}=\frac{1}{r}\\
    \frac{v_y^d}{v_y^u}= \frac{A_x^{u2}-1}{A_x^{u2}-r}\\
    B_x=\mbox{const.}\\
    \frac{B_y^d}{B_y^u}= r \frac{A_x^{u2}-1}{A_x^{u2}-r} \\
    \frac{P^d}{P^u}= 1+ \frac{2}{\beta(1+\tan^2\theta)} \frac{A_x^{u2}}{r}\left[r-1+\frac{r\tan^2\theta}{2A_x^{2u}}\left(1-r^2\left( \frac{A_x^{u2}-1}{A_x^{u2}-r}\right)^2 \right) \right] \label{Eqn:PressureJump}
\end{gather}
where $r$ is the compression across the shock, $A_x^{u2}=\frac{\rho^u v_x^{u2}}{B_x^2}$ is the inflow (upstream) Alfv\'en Mach number squared, $\theta$ is the upstream angle of the magnetic field, and $\beta$ is the upstream plasma-beta.

These ideal MHD jump conditions can be combined with the energy jump relation (Equation \ref{Eqn:EnergyJump}) to be reduced to a single equation\citep{Hau1989} that determines all possible stable shock solutions for a given upstream plasma-$\beta$ and angle of the magnetic field $\theta$:
\begin{gather}
    A_x ^{\text{u}2} = \frac{ A_x ^{\text{d}2} \left( \frac{\gamma-1}{\gamma} \left( \frac{\gamma+1}{\gamma -1} -\tan ^2 \theta \right) \left(A_x ^{\text{d}2} -1 \right) ^2 + \tan ^2 \theta  \left( \frac{\gamma-1}{\gamma} A_x ^{\text{d}2} -1 \right) \left(A_x ^{\text{d}2} -2 \right) \right) - \frac{\beta }{ \cos ^2 \theta } \left( A_x ^{\text{d}2} -1 \right) ^2 } { \frac{\gamma -1}{\gamma} \frac{\left( A_x ^{\text{d}2}-1 \right) ^2}{ \cos ^2 \theta ^{\text{u}}} - A_x ^{\text{d}2} \tan ^2 \theta ^{\text{u}} \left( \frac{\gamma -1}{\gamma} A_x ^{\text{d}2} -1 \right) } \label{eqn:hau}
\end{gather}
These equations are valid for any \textbf{conservative} set of MHD equations, i.e., a set of equations with no sources or sinks of mass/momentum/energy. However, the shock jumps for \textbf{non-conservative} equations are of interest, specifically the optically-thin MHD model, where the radiative losses (and heating term) act as a sink (source) of energy. For such a model, a different approach is required. 

The MHD (no losses) model has neglected the role of viscosity and resistivity in the underlying equations which can lead to heating. However, by defining the upstream and downstream locations to be sufficiently far from the shock, where the gradients of the magnetic and velocity fields are zero, the contribution of these terms vanishes, and the solution reduces to the ideal MHD solution given above. Viscosity and resistivity do however affect the substructure of the shock \citep{Hau1989}.

\subsection{Radiative shock jumps} \label{Sec:RadShock}
\begin{figure}
    \centering
    \includegraphics[width=0.95\linewidth]{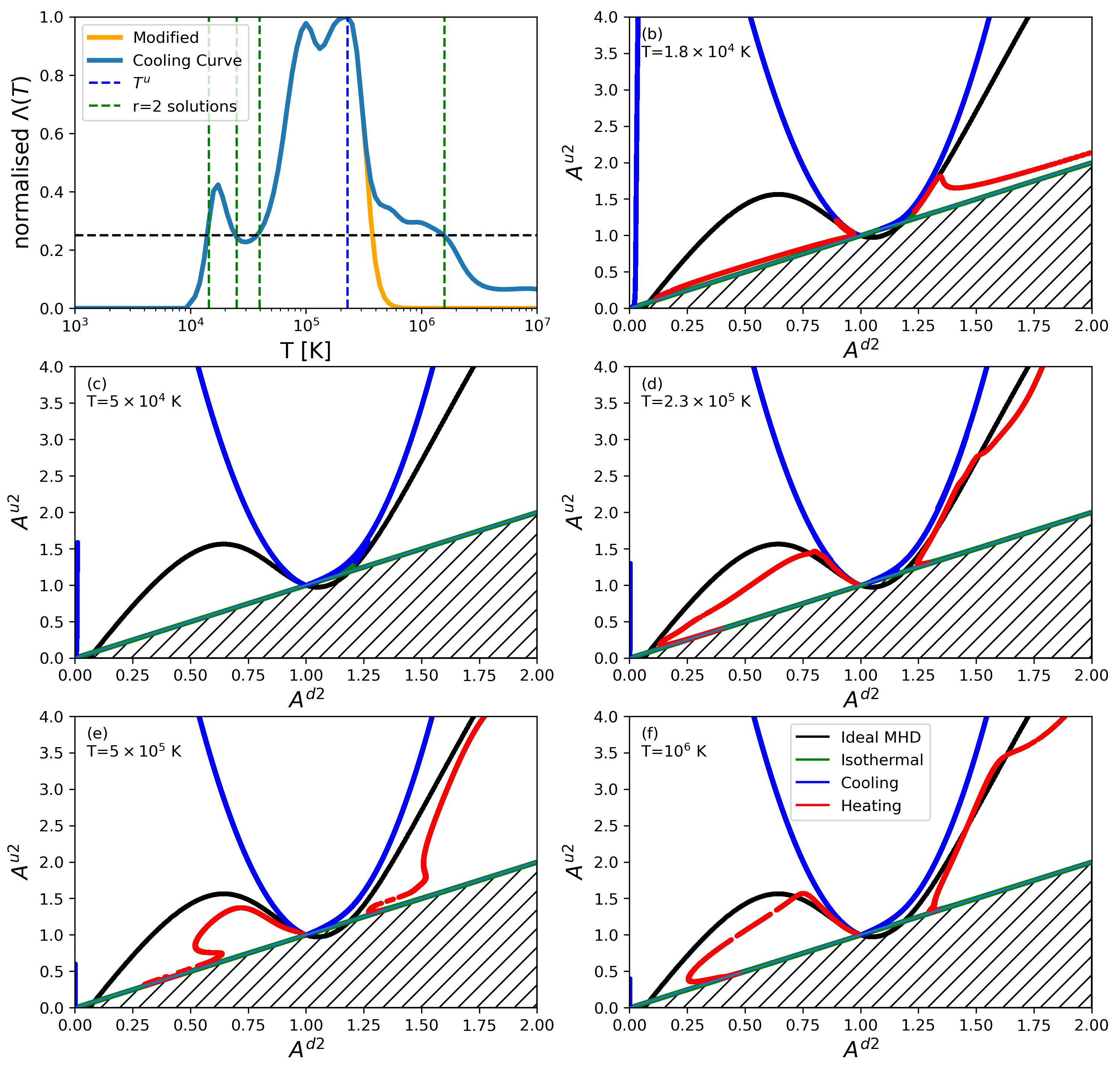}
    \caption{Panel (a) (blue line) Normalised radiative loss curve using CHIANTI v9 \citep{Dere1997,Dere2019} with the default abundance file. The blue and green vertical lines show the upstream and possible downstream temperatures for a shock with $r=2$, as discussed towards the end of \protect\ref{Sec:RadShock}. The orange line is the modified version of the loss curve used in the simulation of Section \protect\ref{Sec:Simulation}. Panels (b)-(e) Example radiative MHD solutions for reference upstream values of  $\beta=0.1,\theta=\pi/8,T^u=10^5$K compared to the MHD solution (black line). The radiative MHD shock solutions are coloured by the temperature jump, with red denoting heating solution, blue denoting cooling solutions, and effectively isothermal solutions in green. Reference upstream temperatures are $T^u=1.8\times 10^4 (b),5.0\times 10^4 (c),2.3\times 10^5 (d),5\times 10^5 (e), 10^6 (f)$}
    \label{fig:ShockCoolingPlot}
\end{figure}

Radiative losses introduce a non-conservative term in the energy equation that fundamentally change the stable conditions across a shock. Applying the same methodology as before (putting the underlying equations in the frame of reference of the shock and assuming a steady-state solution) yields the following: 

\begin{gather}
\nabla \cdot (\rho \textbf{v}) = 0, \\
\nabla \cdot \left( \rho \textbf{v} \textbf{v} + P \textbf{I} - \textbf{B B} + \frac{\textbf{B}^2}{2} \textbf{I} \right) = 0,\\
\nabla \cdot \left[ \textbf{v} ( e + P) -  (\textbf{v} \times \textbf{B}) \times \textbf{B} \right]  =  \rho ^2 \Lambda(T) + \Phi_{H}, \\
\nabla \times (\textbf{v} \times \textbf{B}) = 0, \\
\nabla \cdot \textbf{B} =0.
\end{gather}
The majority of these equations can be directly integrated as before, since there are no sources of mass or momentum, however the energy equation cannot be easily integrated due to the source term. A different approach is required to determine the stable shock jump solutions. 

Following the approach of \citet{Snow2021}, the energy equation can be discarded and replaced with a condition on the equilibrium, namely that for a simple equilibrium, one requires that the heating and the loss terms balance, i.e.,:
\begin{gather}
    \rho ^2 \Lambda(T) - \Phi_{H} = 0
\end{gather}
Such condition must be true sufficiently far from an isolated shock in both the upstream or downstream directions:
\begin{gather}
    \rho^{u2} \Lambda(T^u) - \Phi_{H}^u = \rho^{d2} \Lambda(T^d) - \Phi_{H}^d
\end{gather}
A common approach is to assume a temporally constant heating term to balance the losses at time $t=0$. The heating term is also assumed to be spatially-constant, i.e., $\Phi_H^u=\Phi_H^d$. Under these assumptions, the equilibrium can be simplified as
\begin{gather}
    \frac{\rho^{d2}}{\rho^{u2}}= r^2= \frac{\Lambda(T^u)}{\Lambda(T^d)} \label{eqn:rad_bal}
\end{gather}
i.e., a compression of the medium requires a reduction in the losses. The compression $r$ can also be expressed in terms of the upstream and downstream Alfv\'en Mach number
\begin{gather}
    \frac{A_x^{d2}}{A_x^{u2}}=\frac{1}{r}
\end{gather}
and hence the change in Mach number across the shock also depends on the equilibrium conditions permitted by the loss function. 

The shape of the radiative loss curve varies greatly depending on the medium, composition and physics of interest. Here, as an example, a solar-like optically-thin radiative loss curve is calculated using CHIANTI v9 \citep{Dere1997,Dere2019} with the default abundance file. As a simple example, consider a shock that has a compression of $r=2$ with an upstream temperature at the peak to the radiative loss curve, i.e., $T^u\approx 230,000$ K. From Equation \ref{eqn:rad_bal}, for an equilibrium solution to exist, the losses must decrease by a factor of 4 across the shock. Using the example radiative loss curve in Figure \ref{fig:ShockCoolingPlot}a, this gives 4 potential solutions shown by the intercepts of the horizontal dashed line with the loss curve, which give temperatures of $T^d\approx 14500,25100,39800,1580000$ K. Three of these solutions are potential cooling solutions, where the temperature reduces across the shock. To evaluate if these are shock solutions, these potential solutions need to be analysed using the rest of the shock jump equations.

\subsubsection{Semi-analytic shock solution}

The pressure jump across the shock (Equation \ref{Eqn:PressureJump}) and the ideal gas law $T=\gamma P/\rho$ can be combined to give a temperature jump equation, which, when coupled with the radiative equilibrium condition, yields a pair of equations that can be solved numerically to find the stable shock jump conditions for a given upstream temperature, angle of magnetic field, and plasma-beta:
\begin{gather}
    \frac{T^d}{T^u}= \frac{A^{d2}_x}{A_x^{u2}}\left[ 1+\frac{2}{\beta (1+\tan^2\theta)}\left(A_x^{u2}-A_x^{d2}+\frac{\tan^2\theta}{2} \left(1-\left( \frac{A_x^{u2}-1}{A_x^{d2}-1}\right)^2 \right) \right)  \right] \label{eqn:RMHDSol1}\\
    \frac{\Lambda (T^d)}{\Lambda (T^u)}=\left(\frac{A_x^{d2}}{A_x^{u2}}\right)^2 \label{eqn:RMHDSol2}
\end{gather}
Note that the pressure jump (Equation \ref{Eqn:PressureJump}) arises without using the energy equation, with the pressure term coming from the momentum equation.  The trivial solution where $A_x^{u2}=A_x^{d2}$, i.e., no shock, is satisfied for no jump in temperature $T_u=T^d$, as expected. Other non-trivial solutions exist depending on the shape of the radiative loss curve.

Numerically solving Equations \ref{eqn:RMHDSol1}-\ref{eqn:RMHDSol2} for a choice of upstream parameters as $\beta=0.1,\theta=\pi/8$ for different upstream temperatures is shown in Figure \ref{fig:ShockCoolingPlot}b-e for the CHIANTI cooling curve. Solutions are sought in the range $A^{d2}<2,A^{u2}<4$ with a compression ratio in the range $1.01\leq r \leq 500$. The solution is compared to the ideal MHD solution given by Equation \ref{eqn:hau}. The shaded region of Figures \ref{fig:ShockCoolingPlot}b-f corresponds to a compression $r<1$ and thus does not obey the definition of a shock, thus solutions are not sought in this area. The radiative solution has both heating (red) and cooling (blue) solutions, as well as a few effectively-isothermal shocks where the temperature jump is defined as $0.9<T^d/T^u<1.1$.

The non-trivial ideal MHD solution is a cubic that intersects the linear solution at the slow, Alfv\'en, and fast speeds. The radiative MHD simulation is far less smooth than the ideal solution and cannot be described by a simple (low-order) polynomial. Only solutions with a compression in the range $1.01\leq r \leq 500$ are found by the solver, however this neglects some highly compressible cooling solutions. The CHIANTI cooling curve has zero losses below $T=10^4$ K and thus a cooling solution can be found for any temperature above this, however the compression becomes increasingly large for small $A^{d2}$; Equation \ref{eqn:rad_bal} shows that a big jump in the losses requires a large compression. 

For the 5 tested upstream temperatures, all support cooling shock solutions which follow a roughly cubic shape, see the blue line in Figures \ref{fig:ShockCoolingPlot}b-e. The heating (i.e., temperature increasing solutions) have a far more varied shape. For an upstream temperature of $T^u=1.8\times10^4$K, a heating solution is found that is close to the trivial solution where $A^{u2}=A^{d2}$, i.e., no shock, Figure \ref{fig:ShockCoolingPlot}b. This implies that the heating solutions at this temperature have a small increase in density across the interface. For $T^u=5\times10^4$K, no heating solutions are found, Figure \ref{fig:ShockCoolingPlot}c. The heating solution for reference upstream temperatures of $T=2.3\times10^5$ K and $T=10^6$ are reasonably similar to the MHD solution. Note that this only implies that the compression across the shock is similar, not necessarily temperature. Overall, Figures \ref{fig:ShockCoolingPlot}b-e show the sensitivity of the stable shock solutions relative to the upstream temperature.

The solution in Figures \ref{fig:ShockCoolingPlot}b-e are only valid for the specific reference values of $\beta=0.1,\theta=\pi/8,T^u$. The full parameter space is not presented in this paper and different values of $\beta,\theta,T^u$ will produce different sets of heating and cooling solutions. Appendix \ref{Apx:ExSols} shows a selection of shock solutions for the CHIANTI cooling curve using a fixed temperature and varying the plasma$-\beta$ and $\theta$.

\subsection{Hazy Cooling Curve}

\begin{figure}
    \centering
    \includegraphics[width=0.95\linewidth]{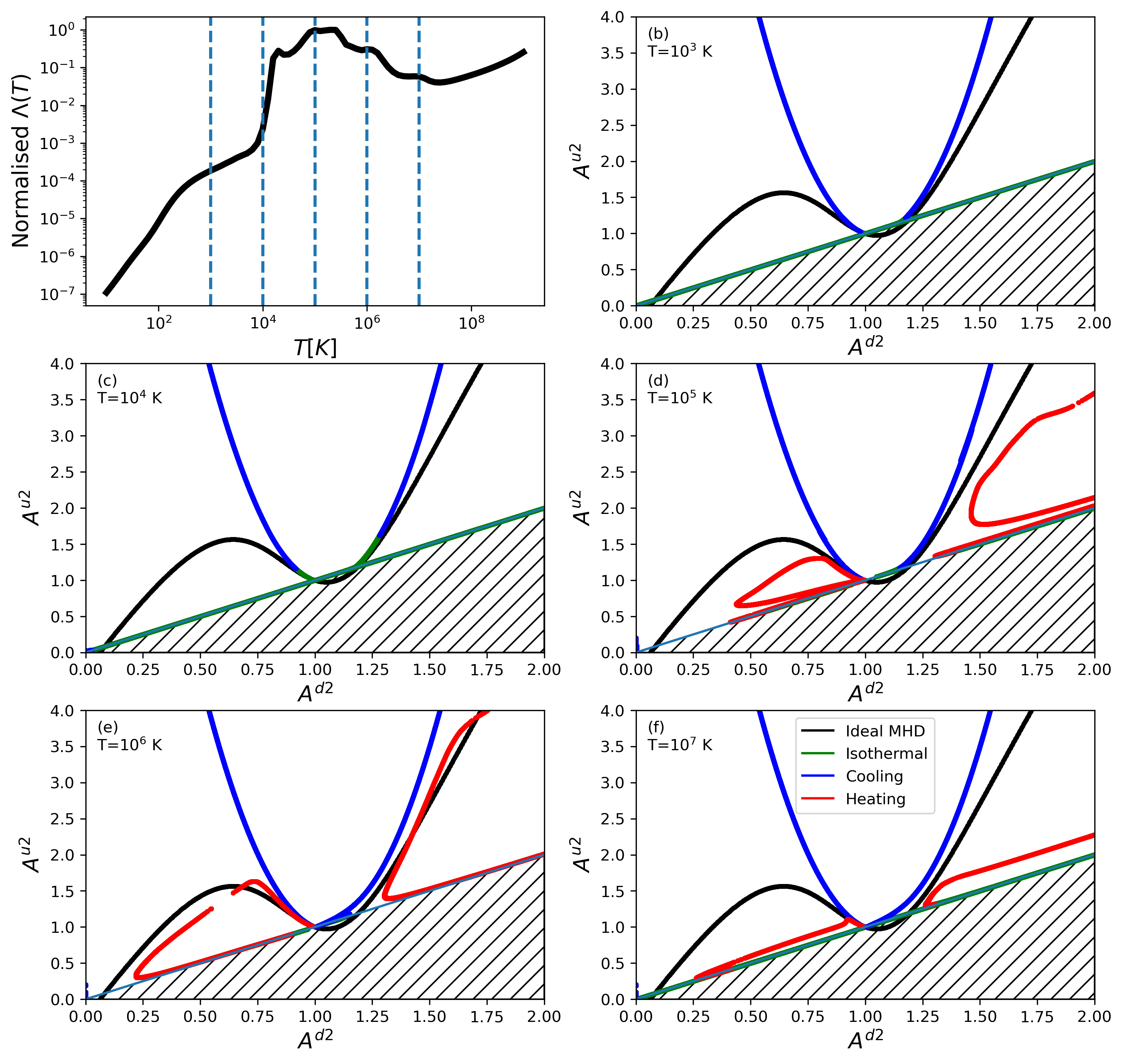}
    \caption{(a) Hazy cooling curve normalised to unity at the peak. The radiative shock solutions are for usptream temperatures of $10^3$K (b), $10^4$K (c), $10^5$K (d), $10^6$K (e), and $10^7$K. The upstream temperatures correspond to the vertical lines in panel (a).}
    \label{fig:ShockCoolingHazy}
\end{figure}

The presence of cooling shock solutions is not unique to the CHIANTI cooling curve used in the previous section. The CHIANTI cooling curve is specific to the solar atmosphere, and other astrophysical systems have different abundances and physical process, hence different temperature dependence of the cooling curve. A common astrophysical loss function is the Hazy cooling curve that is included in the default build of Cloudy version C17 \cite{Ferland2017}, shown in Figure \ref{fig:ShockCoolingHazy}a. The Hazy cooling curve assumes a gas that is in collisional ionisation equilibrium with comic ray heating included for the low temperature regime and solar-like abundances.

As before, Equations \ref{eqn:RMHDSol1}-\ref{eqn:RMHDSol2} can be numerically solved using the Hazy cooling curve as the loss function, and specifying the upstream temperature, plasma-$\beta$ and angle of the magnetic field $\theta$. Again, plasma-$\beta=0.1$ and $\theta=\pi/8$ are held constant and the stable shock solutions are calculated for different upstream temperatures, namely $T^u=10^3,10^4,10^5,10^6,10^7$ K. The radiative shock solutions coloured to indicate the temperature-increasing shocks (red), temperature-decreasing shocks (blue), and essentially-isothermal (green) solutions, shown in Figures \ref{fig:ShockCoolingHazy}b-e.

For upstream temperatures of less than $T^u \approx 10^4$, the radiative loss function is lower than for any hotter temperature and thus only cooling solutions are possible in order to satisfy Equation \ref{eqn:RMHDSol2}, and as such only cooling solutions are found in Figures \ref{fig:ShockCoolingHazy}b,c. Similarly, only cooling solutions are possible above the local loss minimum near the reference temperature of $10^7$ since reductions of the radiative cooling curve only exist for cooler post-shock temperatures. Both heating and cooling solutions are found for the reference upstream temperatures of $T^u=10^5,10^6,10^7$ K, shown in Figures \ref{fig:ShockCoolingHazy}d-f, respectively. As before, the presence of heating or cooling solutions is very sensitive to the upstream temperature. However all tested cases have temperature-reducing shocks as stable solutions of the optically-thin radiative MHD equations. 

\subsection{Limitations of the analytical solutions}

It should be noted that the analytical results presented are for a steady-state, isolated shock and the jump conditions are valid for an undefined distance in the upstream and downstream directions. Substructure may exist between the upstream and downstream states which is not considered in the analysis presented. Also, there is no consideration of the time taken to obtain the equilibrium. For this, numerical simulations are useful in determining the spatiotemporal scale on which cooling solutions can exist. 

The presented shock solutions use a fixed plasma-$\beta$ and $\theta$ value, with the upstream temperature varying. The full parameter space has not been presented in this paper however some additional solutions are shown in Appendix \ref{Apx:ExSols} for the CHIANTI cooling curve.

\section{Numerical Simulation of Tearing Instability} \label{Sec:Simulation}

\begin{figure}
    \centering
    \includegraphics[width=0.95\linewidth]{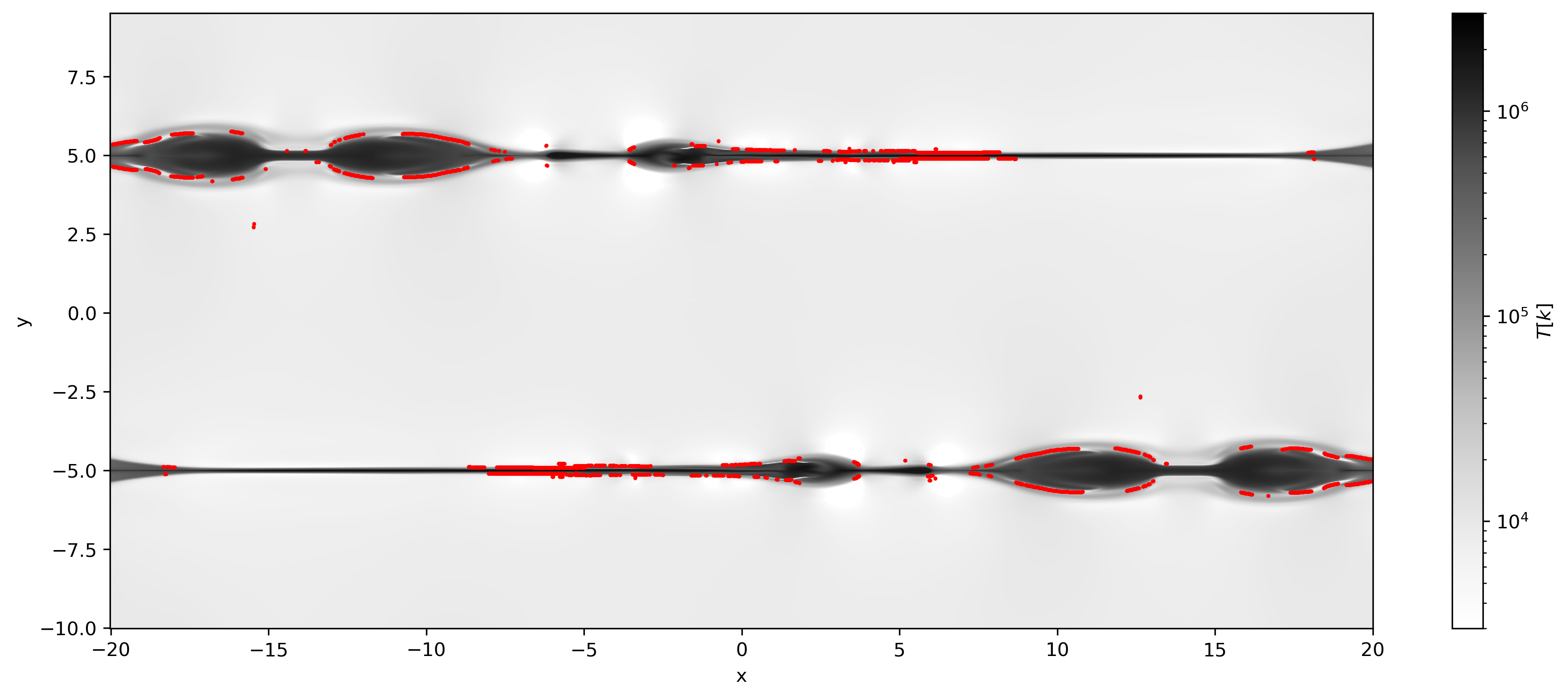}
    \caption{Numerical simulation of the tearing instability at time $t=40$ coloured by the temperature, with the identified slow-mode shocks highlighted in red.}
    \label{fig:reconnectionShocks}
\end{figure}

To investigate the heating and cooling solutions of shock, 2.5D numerical simulations are performed using the (P\underline{I}P) code \citep{Hillier2016} of a tearing-unstable reconnection current sheet with optically-thin radiative losses. Resistivity is included in this model to promote the instability without the reconnection relying on numerical resistivity, specifically, the underlying equations for this simulation are:
\begin{gather}
\frac{\partial \rho}{\partial t} + \nabla \cdot (\rho \textbf{v}) = 0, \\
\frac{\partial}{\partial t} (\rho \textbf{v})+ \nabla \cdot \left( \rho \textbf{v} \textbf{v} + P \textbf{I} - \textbf{B B} + \frac{\textbf{B}^2}{2} \textbf{I} \right) = 0,\\
\frac{\partial}{\partial t} \left( e + \frac{\textbf{B}^2}{2} \right) + \nabla \cdot \left[ \textbf{v} ( e + P) -  (\textbf{v} \times \textbf{B}) \times \textbf{B} +\eta \left(\nabla \times \textbf{B}\right)\times\textbf{B} \right]  =  - \rho ^2 \Lambda(T), \\
\frac{\partial \textbf{B}}{\partial t} - \nabla \times (\textbf{v} \times \textbf{B}-\eta \nabla \times \textbf{B}) = 0, \\
\nabla \cdot \textbf{B} =0, \\
e = \frac{P}{\gamma -1} + \frac{1}{2} \rho v ^2, 
\end{gather}
where the resistivity $\eta=0.001$ is constant and uniform in the entire domain. A modified form of the radiative loss curve is used, where the loss function is smoothly reduced to zero for high temperatures, see Figure \ref{fig:ShockCoolingPlot}a. The gradual decrease in losses was implemented to have a zero-gradient region at extreme temperatures, however this is not a requirement of the code. 

The initial condition is a pair of force-free Harris current sheets with a superimposed perturbation:
\begin{gather}
    B_x=B_0+B_0 \tanh \left( \frac{y-\hat{y}}{w} \right)- B_0 \tanh \left( \frac{y+\hat{y}}{w} \right)+ B_{x,\mbox{pert}}\\
    B_y= B_{y,\mbox{pert}} \\
    B_z=\frac{B_0}{\cosh  \left( \frac{y-\hat{y}}{w} \right)}- \frac{B_0}{\cosh  \left( \frac{y+\hat{y}}{w} \right)}
\end{gather}
where $\hat{y}=5$ is the $y-$location of the current sheet centres, $B_0=1$ is the reference magnetic field strength, and $w=0.1$ defines the width of the current sheet. Since the initial conditions are force-free, the density and pressure (and hence temperature) can be specified as constant, specifically as $\rho=1$, $P=B_0^2\beta/2$, where the plasma-$\beta=0.01$. The pair of current sheets allows periodic boundaries to be specified in all directions, ensuring energy remains within the numerical box, with the exception of the energy lost due to radiation.

The system is normalised according to an initial temperature of $T=10^4$K across the domain. Since the radiative loss curve evaluated at this reference temperature is identically zero, no heating term is required to balance the initial conditions and hence $\Phi_H=0$. The radiative losses are normlised such that the peak value is $\max(\Lambda(T))=0.01$, which, for the reference density of $\rho=1$, means that thermal energy is removed from the system on timescales of 100, which is longer than the simulation time. This is calculated throughout the simulation so that dense regions, such as shocks, feature enhanced losses depending on their local density and temperature.

The reconnection is triggered by specifying a localised pinch in the magnetic field
\begin{gather}
    B_{x,\mbox{pert}}=B_{0,\mbox{pert}}\frac{x}{2}e^{-\frac{(x-\hat{x})^2+(y-\hat{y})^2}{w/16}}-B_{0,\mbox{pert}}\frac{x}{2}e^{-\frac{(x+\hat{x})^2+(y+\hat{y})^2}{w/16}}  \\
    B_{y,\mbox{pert}}= B_{0,\mbox{pert}}\frac{y}{2}e^{-\frac{(x-\hat{x})^2+(y-\hat{y})^2}{w/16}}-B_{0,\mbox{pert}}\frac{y}{2}e^{-\frac{(x+\hat{x})^2+(y+\hat{y})^2}{w/16}}
\end{gather}
with reference amplitude $B_{0,\mbox{pert}}=0.1$ which is a small perturbation centred on $\pm[\hat{x},\hat{y}]$ that decays with distance. The distance from the current sheets is chosen to limit the interaction between the reconnection regions on the timescale of the simulation. The simulation is resolved using $4096\times2048$ grid cells spanning $-20\leq x\leq 20$, $-10\leq y\leq 10$ ($dx=dy\approx0.01$) with a 1st order HLLD solver.

As the initial conditions evolve, the current sheet reconnects and becomes tearing-unstable leading to a wealth of smaller plasmoids being generated, as shown by the snapshot at time $t=40$ in Figure \ref{fig:reconnectionShocks}. The evolution of the tearing instability is well discussed in the literature \citep{Zanna2016,Loureiro2016}. Here, the purpose is not to study the intricacies of the tearing instability, but rather use is as a test bed for identifying and studying shocks. Plasmoids generated by fast magnetic reconnection are known to host a wealth of shocks \citep{Zenitani2015} making this simulation ideal for identifying the presence of heating and cooling shocks.

The system features a broad distribution of temperature that spans the full extent of the cooling curve, as shown in Figure \ref{fig:reconnectionShocks}. The background temperature remains stable at $T=10^4$, however the reconnection region features strong increases to a few million Kelvin. The high temperatures produced are a result of the low plasma beta used in the simulation, which provides a large reservoir of magnetic energy that can be converted to thermal and kinetic energy during magnetic reconnection.

\begin{figure}
    \centering
    \includegraphics[width=0.95\linewidth]{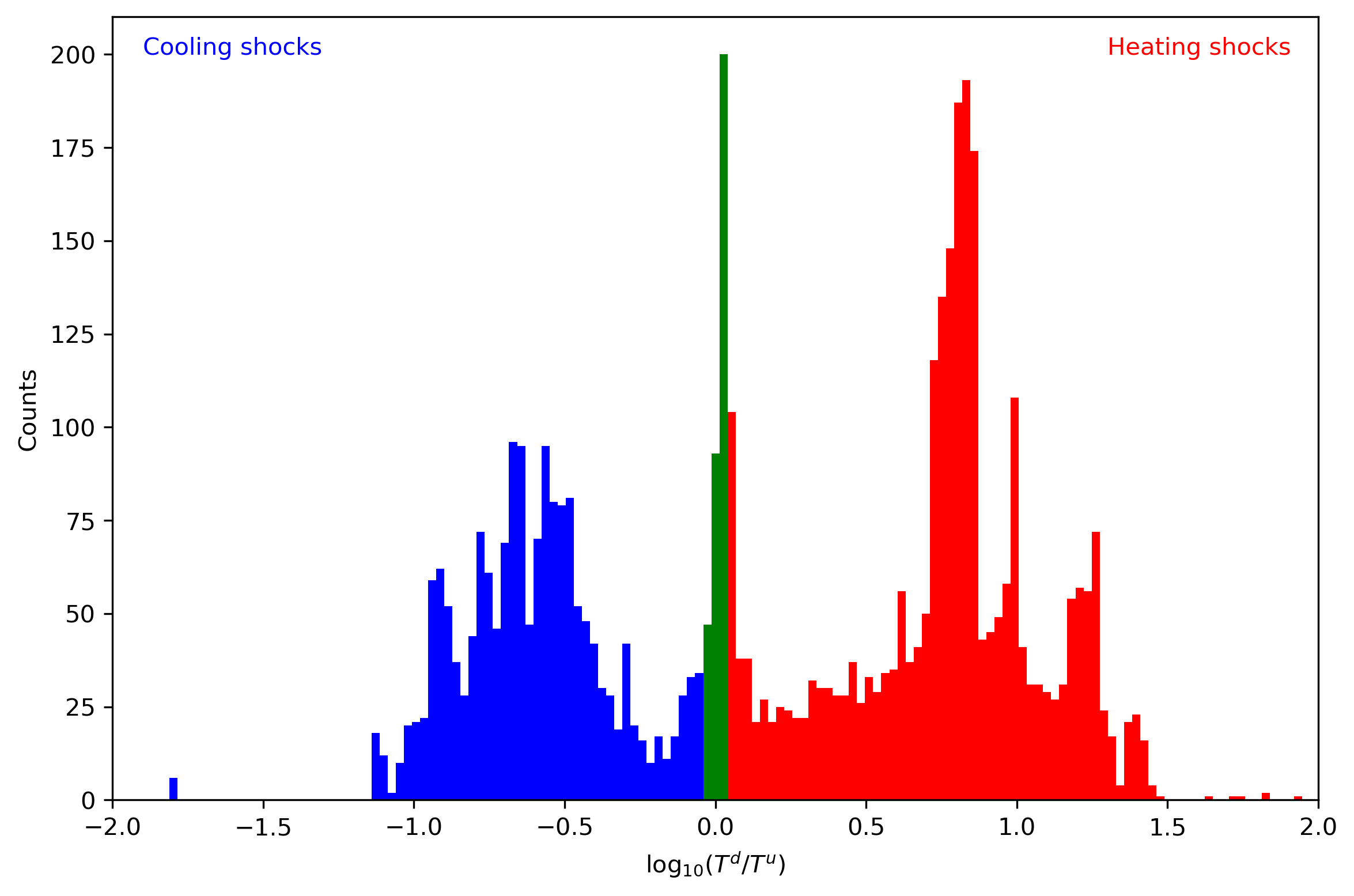}
    \caption{Histogram of the temperature jump of the detected slow-mode shocks in the tearing instability coloured.}
    \label{fig:shockTemperature}
\end{figure}

\subsection{Shock Detection}

Shocks are detected and classified using the open-source ShockID routine \citep{Snow2021ShockID} which has recently been ported to Python with additional features such as parallisation. The shock detection method works by identifying candidate shocks from locations where $\nabla \cdot \textbf{v}<0$, i.e., converging flow, calculating the shock direction from the maximum density gradient, interpolating the velocity, magnetic field, density, and pressure along the shock direction, and finally using these values to classify the shock transition. The result is a list of grid cell locations that satisfy shock jump transitions. Figure \ref{fig:reconnectionShocks} shows the detected slow-mode shocks at an instance in time. It should be noted that the shock detection method is approximate and only detects whether the candidate cell satisfies the shock jump equations.

The majority of the detected shocks are categorised as slow-mode shocks and centred around the plasmoids, which is to be expected from the known structure of shocks around plasmoids \citep{Zenitani2015}. Here only the slow-mode shocks are analysed in detail since these are the most numerous in the simulation. 

\begin{figure}
    \centering
    \includegraphics[width=0.95\linewidth]{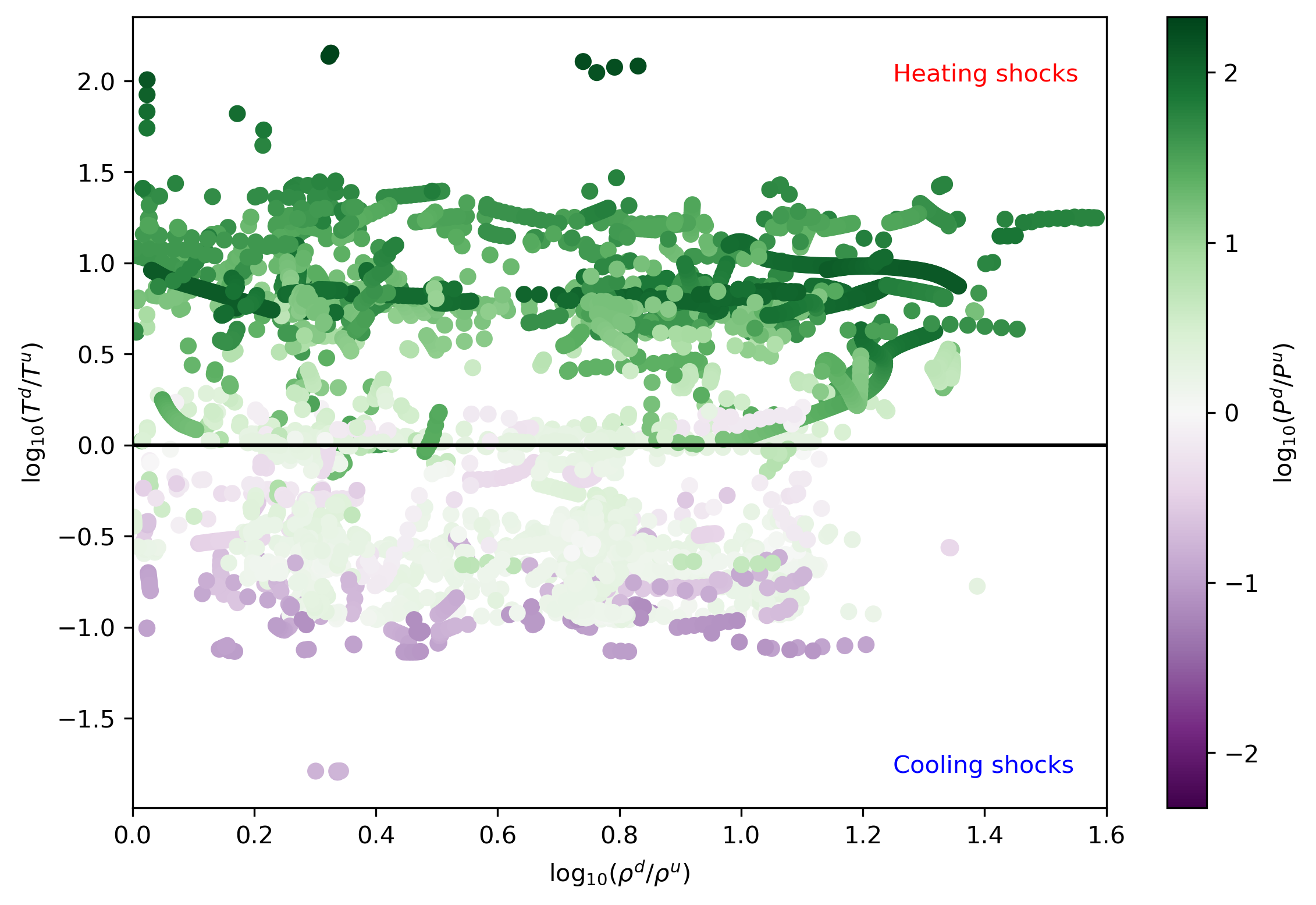}
    \caption{Scatter plot of the compression vs the temperature jump for the detected slow-mode shocks, coloured by the pressure jump}
    \label{fig:shockPressure}
\end{figure}

\begin{figure}
    \centering
    \includegraphics[width=0.95\linewidth]{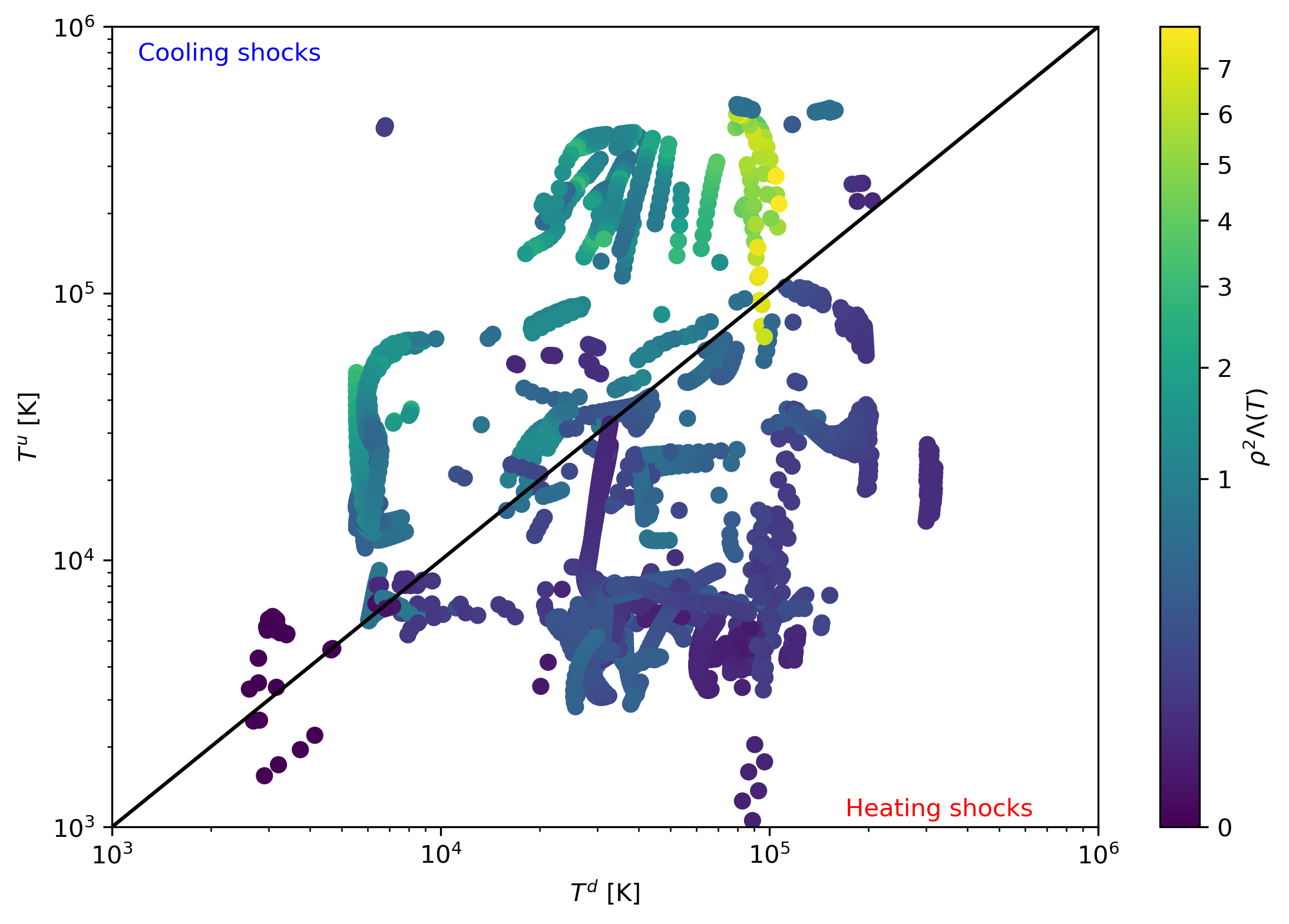}
    \caption{Scatter plot of the upstream temperature vs the downstream temperature for the detected slow-mode shocks, coloured by the total losses along the 1D shock width. The black line represents an isothermal temperature across the shock.}
    \label{fig:shockLoss}
\end{figure}

A histogram of the temperature jumps across the shocks is shown in Figure \ref{fig:shockTemperature}, where 'across the shock' is a distance of fewer than 5 grid cells in the upstream and downstream direction (the exact length of the shock depends on the density profile along the shock line and the maximum line length checked is 5 cells either side of the shock for a total of 11 grid cells). Three broad categories of shocks exist: heating shocks, cooling shocks, and near-isothermal shocks. The near-isothermal shocks are determined as having a temperature jump $0.9<=T^d/T^u<=1.1$. The temperature jump across the shock is calculated local to the shock, as in within 5 grid cells in the upstream and downstream direction. There are 2670 heating solutions, 1722 cooling solutions, and 357 near-isothermal shocks detected in the simulation. Heating and cooling solutions are therefore abundant in the simulation. This can be seen by cool `halos' that form around the plasmoids in Figure \ref{fig:reconnectionShocks}.

The detected compression across the shocks is in the range $1<r<40$, with a large range of compression in both the heating and cooling shocks, see Figure \ref{fig:shockPressure}. In MHD, it is possible to determine a maximum compression ratio as a function of the adiabatic index by considering an infinitely fast fast-mode shock, which yeilds:
\begin{gather}
    r_{max}=\frac{\gamma+1}{\gamma-1}
\end{gather}
which, for $\gamma=5/3$ which is used in this paper, yields a maximum compression of $r_{max}=4$. In the radiative simulation, slow-mode shocks are detected with an order of magnitude more compression than is possible in ideal MHD. The enhanced compression is seen in other simulations of shocks with non-conservative energy equations to account for losses of macroscopic fluid energy \citep{Snow2021,Snow2023}.

The detected heating shocks are generally associated with a increase in pressure across the shock, as shown in Figure \ref{fig:shockPressure}. This is similar to MHD, where a pressure jump is necessary for a shock. However, for the cooling shocks, there are cases where the pressure decreases and increases. Cooling shocks can feature a decrease in temperature despite an increase in pressure, as long as the pressure increase is small compared to the density increase. Similarly isothermal shock solutions can be obtained when the density and pressure jumps are roughly equal.

A cooling shock that has a pressure jump less than unity presents an odd situation where the pressure and density gradients act in opposite direction. As such, shocks of this nature may become baroclinically unstable, however, this is not detected in the simulation. Further study is needed on the long-term stability of the cooling shocks with pressure reductions.

Figure \ref{fig:shockLoss} shows a scatter-plot of the upstream and downstream temperatures coloured by the total losses along the 1D shock width. Both heating and cooling shocks exist across a wide range of temperatures in the upstream and downstream regions. This was expected from Section \ref{Sec:RadShock} where a single upstream temperature allows stable solution to exist for multiple downstream temperatures, which can be either heating or cooling. The total losses along the shock, calculated as 
\begin{gather}
    L=\sum^{u}_{d} \rho^2 \Lambda(T)
\end{gather}
gives the losses per unit area over the finite width of the shock. A trend exist whereby the total losses are stronger for cooling solutions. This is reasonably intuitive since for a cooling shock to exist, the energy lost due to radiation must exceed the adiabatic temperature increase due to compression. The timescale for the losses becomes significantly enhanced in all the shocks due to the large increases in density across the interface. For the initial reference density, losses occur on timescales of $10^{2}$ at the peak of the cooling curve. Here, losses are seen to have timescales of around $10^{-1}$, three orders of magnitude quicker than the reference values. As such, radiative losses are very efficient in the shocks.


\section{Discussion}

This paper has proven both analytically and numerically that the optically-thin radiative MHD equations can support both heating and cooling shock solutions. The presence of cooling solutions depends on the radiative cooling curve and the upstream temperature. 

\subsection{Model assumptions}

The underlying model used for this analysis is widely used across a number of astrophysical plasmas. The results of this paper are only as valid as the assumptions of the underlying model, as is true of any analytical or numerical study.  

A strong assumption here is that the heating function is constant. In \citet{Snow2021}, a related set of equations was studied where the heating function was constant and energy loss was due to the ionisation energy required to change the state of the atom. It was concluded that their underlying model necessarily resulted in shocks that reduce the temperature of the plasma. However, in a subsequent paper, the underlying model was extended to include more realistic treatment of hydrogen and a time dependent heating function revealed that the same type of shock (switch-off slow-mode) now featured an increase in temperature postshock \cite{Snow2023}.

The heating function should likely be dependent on the local properties of the plasma, e.g., for photon heating, the energy absorbed by the plasma is dependent on the optical depth and hence the density. At shock fronts, the density increases greatly (up to a factor of 40 was found in the simulation) and hence one may expect that the energy absorbed by photons would increase, leading to locally enhanced heating in the post-shock region. Further work is required to study the consequences of non-constant heating functions and how this affects the post-shock temperature. 

The analytical solutions assume steady-state shocks and only consider the jump conditions an arbitrary distance from the shock. In dynamic systems, the time taken to obtain a steady-state shock may be long and thus shocks in observations and simulations may not strictly adhere to the developed solutions. However, the simulation presented in this paper shows that the cooling solution can be obtained over a relatively short distance and hence provides evidence of cooling shocks forming on small lengthscales.

\subsection{Interstellar medium}

Shocks in the interstellar medium are often studied using the isothermal assumption since the losses within the shock are assumed to be strong enough to balance the compressional temperature increase post-shock. This has been used to explain the observed presence of molecules in the post-shock regions that should have been disassociated by the MHD temperature predictions \citep{Draine1993}. As such, a modelling approximation is to the set the adiabatic index to unity such that the system is isothermal.

For typical temperatures of the interstellar medium ($T<10^4$ K), the stable shock solutions using the Hazy cooling curve result in only cooling solutions, see Figure \ref{fig:ShockCoolingHazy}. Cooling shocks could also explain the observed lack of molecular disassociation. However, the cooling shock solutions are not possible to obtain by changing the adiabative index alone, since this would require $\gamma <1$ local to the shock. Setting $\gamma <1$ is problematic from a thermodynamic standpoint. The reason why such departures from classical thermodynamic laws occur for this model is that the problem becomes non-local, with energy available to leave the system through the radiative losses.

\subsection{Relation to the thermal stability}

The thermal instability occurs when the radiative losses cause a reduction in the pressure, driving an inflow which enhances the temperature, allowing for additional cooling. The thermal instability is critical for the formation of cool condensations in the solar atmosphere\cite{Field1965,Sen2022}, planetary nebulae\cite{Hunter1970}, and the intergalactic medium \cite{Sanchez2002}. It is possible that non-steady cooling shocks could result in a post-shock condition that is unstable to the thermal instability, potentially enhancing the formations of condensations.

The idea of shock-induced coronal condensations has been proposed in other papers that use very similar underlying model equations, i.e., MHD with optically-thin radiative losses \citep{Forbes1991}. 
The post-shock temperature calculated in \citet{Forbes1991} is based on the MHD predicted model and thus shocks were assumed to heat, then gradually cool post-shock over relatively long timescales. In this paper it has been demonstrated that cooling solutions are possible and can occur over short length-scales. As such, cooling shock solutions may be an efficient way to induce condensations, which requires further study.

\section{Conclusion}

In this paper, it has been shown both analytically and numerically that the radiative MHD equations allow both heating and cooling solutions of shocks to exist, where a cooling shock has a temperature decrease across the interface. Introducing optically-thin radiative losses results in a non-conservative energy equation, however the shock jump conditions can be closed through consideration of an equilibrium state either side of the shock, resulting in a pair of equations that can be numerically solved to find the steady-state shock solutions for given upstream angle of magnetic field, plasma beta, and temperature.

The analytical solution shows that radiative shocks have a strong departure from the ideal MHD solution and multiple shock solutions exist that can be heating or cooling. However, this is under the assumption that the shocks are steady state and the jump conditions have reached an equilibrium over some unknown distance. In the simulation data, the same behaviour is found, i.e., both heating and cooling solutions, when analysed over a finite and small distance specified as 5 grid cells in the upstream and downstream direction. This shows that cooling shocks can be obtained locally. Since the optically-thin radiative losses considered in this paper are dependent on the density squared, compression at the shock front can drastically increase rate at which radiative losses remove energy. 

The shape of the cooling curve is important in determining the presence of heating and cooling solutions. Equation \ref{eqn:rad_bal} relates the loss rate either side of the shock to the compression:
\begin{gather}
    \frac{\rho^{d2}}{\rho^{u2}}= r^2= \frac{\Lambda(T^u)}{\Lambda(T^d)}
\end{gather}
For a hypothetical radiative loss curve that monotonically decreases with temperature, a compression of the medium requires a temperature increase across the interface. Similarly, a monotonically increasing loss curve would only allow cooling shock solutions. Realistic radiative loss curves often have a complex shape with multiple turning points and thus allow both heating and cooling solutions to exist.

Only the optically-thin regime is studied here. For optically-thick plasmas, the emitted photons can become reabsorbed and potentially lead to heating processes. This process may become important for shocks, where the density increases greatly, and thus the opacity may increase. As such, including radiative heating processes in the underlying equations may provide extra heating or even inhibit cooling shocks from forming. Further work is required to study this effect.

Often, shocks are studied for their role in heating. However, for the optically-thin raditive MHD model presented here, shocks can also provide substantial cooling due to the locally enhanced radiative losses in the finite-width of the shock. As such, the thermal contribution of shocks is far less definitive when radiative losses are included than ideal MHD results would imply. 

\begin{acknowledgments}
BS is supported by STFC research grant ST/V000659/1. 
I would thank Prof. Andrew Hillier (University of Exeter) and Dr. Malcolm Druett (KU Leuven) for support and feedback with this research. 
\end{acknowledgments}

\section*{Data Availability Statement}

The data that support the findings of this study are available from the BS upon reasonable request. The (P\underline{I}P) solver is freely available on GitHub: \url{https://github.com/AstroSnow/PIP}; as is the ShockID code: \url{https://github.com/AstroSnow/shockid}. The post-processing scripts, including the codes required to find shock solutions for given upstream properties with a user-specified loss function, are available here: \url{https://github.com/AstroSnow/RadiativeShockCool}.

\appendix

\section{Shock solutions for different plasma$-\beta$ and $\theta$} \label{Apx:ExSols}

\begin{figure}
    \centering
    \includegraphics[width=0.9\textwidth]{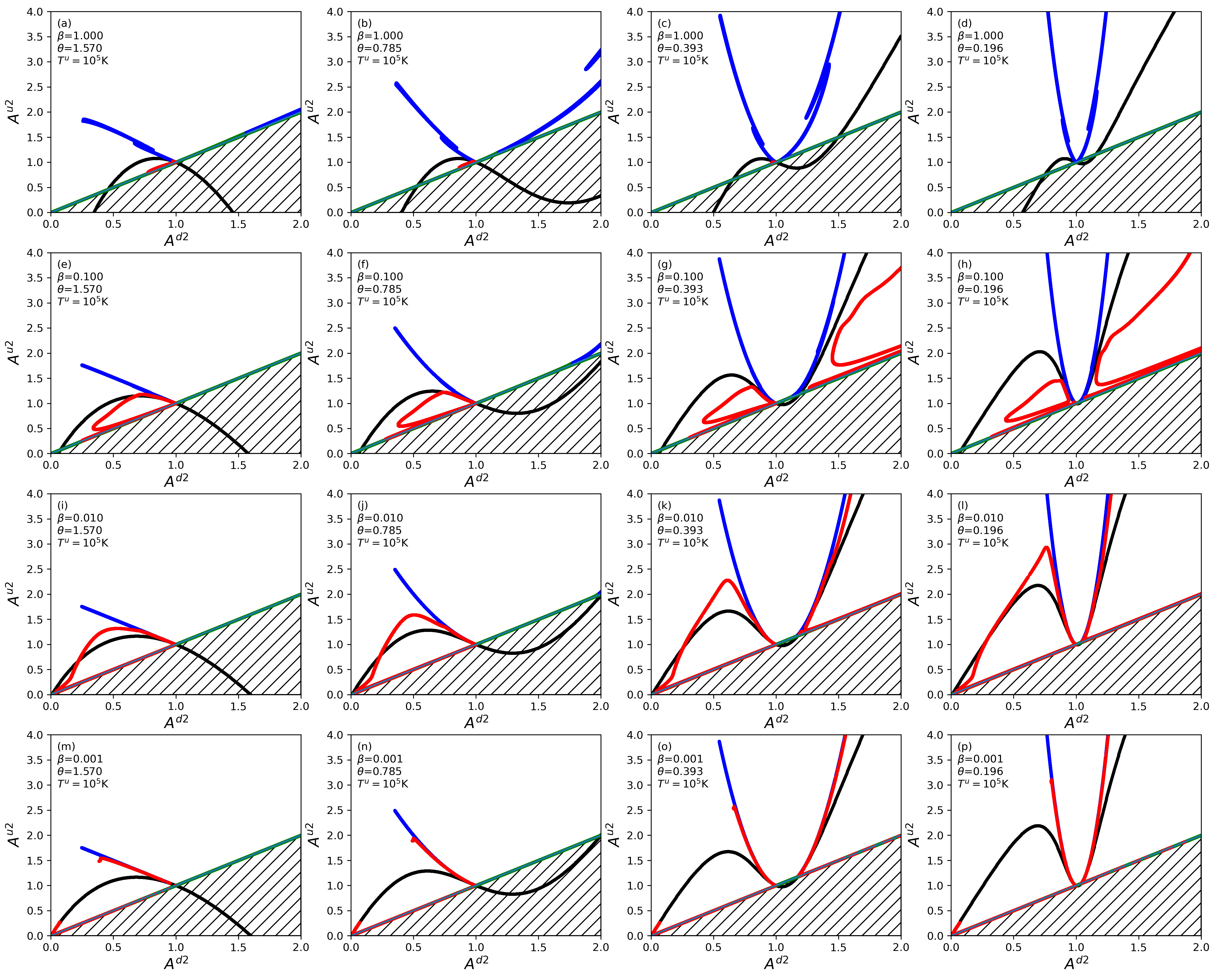}
    \caption{Shock solutions for the CHIANTI cooling curve with a constant upstream temperature of $T^u=10^5$K and varying plasma-$\beta$ and $\theta$ values. Cooling solutions are shown in blue, and heating solutions in red.}
    \label{fig:chianti_param}
\end{figure}

Figures \ref{fig:chianti_param} shows shock solutions for the CHIANTI cooling curve using different plasma-$\beta$ and $\theta$ values for a constant upstream temperature of $T^u=10^5$K. Solutions are sought in the compression range $1.01 \leq r \leq 50$ for downstream Alfv\'en Mach numbers of $0<A^{d2}<2$.

\bibliography{RadShockCool}

\end{document}